\documentclass[%
 reprint,
 superscriptaddress,
 groupedaddress,
 unsortedaddress,
 runinaddress,
 frontmatterverbose, 
 showpacs,preprintnumbers,
 nofootinbib,
 nobibnotes,
 amsmath,amssymb,
 aps,
 pra,
 prb,
 rmp,
 prstab,
 prstper,
 floatfix,
 ]{revtex4-1}
\usepackage{amsmath}
\usepackage{amssymb}
\usepackage{units}
\usepackage{color}
\usepackage{graphicx}
\usepackage{dcolumn}
\usepackage{bm}
\usepackage{fixmath}
\usepackage{float}
\usepackage{upgreek}

\def\lsop#1{{\boldsymbol{\mathcal{#1}}}}
\def\lsopBar#1{{\bar{\boldsymbol{\mathcal{#1}}}}}
\def\lop#1{\hat{\boldsymbol{#1}}}
\def\lopBar#1{\hat{\boldsymbol{\bar{#1}}}}
\begin{document}


\title{Exciton Transfer Using Rates Extracted From the ``Hierarchical Equations of Motion''}

\author{Joachim Seibt}
\email{joachim.seibt@uni-rostock.de}
%
%
%
\author{Oliver K\"{u}hn}%
\affiliation{%
Institute of Physics, University of Rostock, Albert-Einstein-Str. 23-24, 18059 Rostock, Germany
}%
%

\begin{abstract}
Frenkel exciton  population dynamics of an excitonic dimer is studied by comparing results from a quantum master equation (QME) involving rates from second-order perturbative treatment with respect to the excitonic coupling with non-perturbative results from ``Hierarchical Equations of Motion'' (HEOM).   By formulating generic Liouville-space expressions for the rates, we can choose to evaluate them either via HEOM propagations or by applying cumulant expansion.   The coupling of electronic transitions to bath modes is modeled either as overdamped oscillators for description of thermal bath components or as underdamped oscillators to account for intramolecular vibrations. Cases  of initial nonequilibrium and equilibrium vibrations are discussed. In case of HEOM  initial equilibration enters via a polaron transformation. 
Pointing out the differences between the nonequilibrium and equilibrium approach in the context of the projection operator formalism, we identify a further description, where the transfer dynamics is driven only by fluctuations without involvement of dissipation. Despite this approximation, also this approach can yield meaningful results in certain parameter regimes.
While for the chosen model HEOM has no technical advantage for evaluation of the rate expressions compared to cumulant expansion, there are situations where only evaluation with HEOM is applicable. For instance, a separation of  reference and interaction Hamiltonian via a polaron transformation to account for the interplay between Coulomb coupling and vibrational oscillations of the bath at the level of a second-order treatment can be adjusted for a treatment with HEOM.
\end{abstract} 

\maketitle


\section{Introduction}

The dissipative excited state dynamics  of molecular aggregates is often treated by separating the relevant system from an environment \cite{May11,VaAbMa13}. The latter is usually identified with a thermal bath of low-frequency modes of the surrounding molecular matrix, such as a protein scaffold \cite{kuhn97_4154,renger02_9997,CuYaWe20_JPCB}, but can also comprise intramolecular vibrational modes of monomers in the aggregate \cite{renger01_137,AdRe06_BPJ_2778,LiZhBa14_JCP_134106,nalbach15_022706,liu16_272,SoMaLi19_PRL_100502}. 
In the ``Hierarchical Equations of Motion'' (HEOM) method all nuclear degrees of freedom are attributed to the environment \cite{Ta12_JCP_22A550}, and all orders of perturbation theory with respect to their interaction
with the system built from purely electronic components are taken into account at least in principle~\cite{Tanimura06JPCJ}. Although it yields accurate results, due to the complexity of the calculation it is not immediately possible to get insight into mechanistic aspects of transfer processes. For treatment at a lower  level of accuracy often second-order rate theories are used (for a consideration of higher-order effects, see Ref. \citenum{may09_10086}.), which are applicable in cases where a parameter relevant for the transfer process is sufficiently small to be treated in a perturbative way. In the case of the F\"{o}rster approach formulated in the localized basis with assignment of electronic excitations to monomer units the excitonic coupling is taken into account to second order, whereas in the Redfield approach formulated in the exciton basis the second-order perturbation is taken with respect to the system-bath coupling \cite{YaFl02_CP_355}. In both approaches the dynamics of populations and coherences is usually separated in the framework of the secular approximation. Furthermore, the Markov approximation of time-local treatment enters \cite{May11}. 
While in the standard formulation of the F\"{o}rster approach and of the so-called modified Redfield approach (but not of the standard Redfield approach) a thermal equilibration in the electronically excited initial state of the transfer process is assumed, also a non-equilibrium formulation is possible to account for the situation that in general the environment is not thermally equilibrated after electronic excitations \cite{kuhn96_5821,SeMa17_JCP_174109,DiRe16_JCP_034105,Trushechkin19_JCP_074101}. 

Recently, the role of intramolecular vibrations, often modeled as underdamped oscillators, in excitation transfer processes has been discussed in the literature from different perspectives
\cite{christensson12_7449,polyutov12_21,AbVa16_PhotosynthRes_33,nalbach15_022706,liu16_272,DiWaCa15_JPCL_627,JuCs18_AIP_Adv_045318,KiLeKi9_JPCA_1186,schroter15_536,LeTr17_JCP_075101,MaSoNo16_ChemPhysChem_1356,OROl14_NatComm_3012,WaZh19_JCompChem_1097}. 
Vibrations with description by, e.g.\ a Brownian oscillator model, can be treated in the framework of F\"{o}rster theory within a limited regime defined by the ratio of excitonic coupling and reorganization energy of the vibration. Moreover, the question may arise to which extent the F\"{o}rster approach is applicable outside its actual range of validity when vibrations become involved.
There are also ways to extend the usual second-order treatment in the localized basis beyond the concept of treating only the excitonic coupling as such perturbatively to second order \cite{JaChReEa08_JCP_101104}. 
As the second-order transfer rates can be formulated in a generic way in terms of Liouville-space expressions~\cite{yan88_4842,kuhn96_5821}, it should be possible to calculate these rates directly via HEOM propagations. This would allow for flexible adjustments of the rate calculations without methodological changes or cumbersome derivations of an analytical form of the specific rate expressions.

In the present work we are aiming at a comprehensive understanding of usually disregarded aspects of the widely used second-order rate theories. We concentrate on the description in the localized basis 
with F\"{o}rster-type interaction Hamiltonian, however, going beyond the usual assumptions entering in F\"{o}rster treatment of population transfer. The population dynamics will be described by a quantum master equation (QME) involving transfer rates obtained from different approaches of evaluating the generic Liouville-space rate expressions. Specifically, we will contrast rates obtained either via HEOM propagation or by applying cumulant expansion to the rate expressions in terms of line shape functions. Comparing the resulting dynamics with numerically exact HEOM reference data, we can scrutinize  the influence of various aspects of the rate calculation.  In particular, we investigate the role of initial thermal equilibration and the effect of performing a back-transformation from the interaction to the Schr\"odinger representation.

By tuning the excitonic coupling and the energy gap between the electronic excitation energies of the monomer units, we study the role of over - and underdamped oscillators in the transfer process.
To study how the assumption of thermal equilibration in the excited initial state of the transfer process influences the population dynamics, we apply the concept of polaron transformation in the context of the HEOM formalism for the first time.
Furthermore, we outline an approach for second-order perturbative treatment with separation of reference and interaction Hamiltonian via a polaron transformation. Such a separation is already known from the literature, but has only been formulated for a thermal phonon bath so far and not for underdamped oscillators~\cite{SiHa84_JCP_2615,LeMoCa12_JCP_204120,XuWaYaCa16_NJP_2016,Ki16_JCP_123,Ki16_JCP_70}. 

This article is organized as follows: In Section \ref{sec:theoretical_background} we introduce the theoretical background, including general aspects of the HEOM approach, extraction of rates, polaron transformation and specification of rate expressions in terms of line shape functions under different assumptions. In section \ref{sec:results} we present numerical results, thereby addressing the different aspects mentioned above. Advantages of using HEOM for the calculation of second-order rates are discussed in Section \ref{sec:discussion}.  In Section \ref{sec:conclusions} we draw conclusions from the present findings and outline concepts for further investigations in forthcoming works.
%
\section{Theoretical background} \label{sec:theoretical_background}
\subsection{Exciton-Vibrational Hamiltonian}
%
The exciton-vibrational Hamiltonian to be used in the following is rather standard and of the general system-bath form
$\hat{H}=\hat{H}_{\rm S}+\hat{H}_{\rm B}+\hat{H}_{\rm SB}$~\cite{May11,schroter15_1}. 
The (excitonic) system part is given by 
\begin{equation} \label{eq:Hamiltonian_general_system}
\hat{H}_{\rm S}=\sum_{mn} ( \delta_{mn} (\epsilon_m + \lambda_m) + J_{mn} )\hat{B}^{\dagger}_m \hat{B}_n \,. 
\end{equation}
Here, $\epsilon_m$ and $J_{mn}$ are site energies and Coulomb coupling, respectively, and we further introduced the exciton creation, $\hat{B}^{\dagger}_m$, and annihilation, $ \hat{B}_m$, operators. The state $m$ implies that the $m$ monomer is in the electronically excited state $e_m$, whereas all other monomers are in the electric ground state $g_n$.  

Two different models concerning the system-bath, i.e. exciton-vibrational, coupling will be considered (for a general classification, see Ref. \cite{schroter15_1}).
In case of a thermal bath described by (phonon) coordinates $\{x_i\}$ and frequencies  $\{\omega_i\}$ we have ($\hbar=1$)
\begin{equation}
  \label{eq:Hamiltonian_general_bath}
\hat{H}_{\mathrm B} = \sum_i\frac{\omega_{i}}{2} \Big(-\frac{\partial^2}{\partial x_{i}^2} + \hat x_{i}^2 \Big    ) \, .
\end{equation}
The coupling of phonon modes to electronic transitions is commonly described by the displaced oscillator model with the Hamiltonian
\begin{align}
\label{eq:HSB1}
 \hat{H}_{\rm SB}& = - \sum_{m,i} \omega_{i}   g_{m,i} \hat x_i \hat{B}^{\dagger}_m \hat{B}_m \nonumber\\
 &=\sum_{m} \hat u_{m}  \hat{B}^{\dagger}_m \hat{B}_m 
\end{align}
where we introduced the gap coordinate operator $\hat u_m$. For this model the total reorganization energy at monomer $m$ entering Eq. \eqref{eq:Hamiltonian_general_system} is given by $\lambda_m=\sum_i \omega_i g_{m,i}^2/2$.

 The actual distribution of couplings $\{g_{m,i}\}$ will be described by a Debye-Drude (DD) spectral density, which will be taken for all sites (skipping the site index) 
\begin{equation} \label{eq:Debye-Drude_spectral_density}
J_{\rm DD}(\omega)= 2 \lambda_{\rm DD} \frac{\omega_{\rm c} \omega}{\omega^2+\omega_{\rm c}^2}
\end{equation}
Here, $1/\omega_{\rm c}$ and  $\lambda_{\rm DD}$  are bath correlation time and  bath reorganization energy upon electronic excitation, respectively~\cite{Mukamel95}.

In the second system bath model, we will describe the case of coupling to a damped high-frequency (intramolecular) mode. This is commonly done using the multimode Brownian oscillator (BO) model. Here, $\hat{H}_{\mathrm B}$ and 
$ \hat{H}_{\rm S-B}$ are combined as follows:
\begin{align}\label{eq:MBO}
	 \hat { H}_{\rm BO}^{} &\equiv  \hat{ H}_{\rm B} +  \hat{ H}_{\rm SB}\nonumber\\ 
	 &=\sum_i\frac{\omega_{i}}{2} \Big[-\frac{\partial^2}{\partial x_{i}^2} + \Big(x_{i} - \sum_{m,\xi} \frac{c_{\xi,i}^{(m)}
     { q}_{m,\xi}}{\omega_i}
 \Big)^2 \Big] 
\end{align}
where $\{q_{m,\xi}\}$ denotes the set of intramolecular  modes with frequencies $\{\omega_{m,\xi}\}$ for site $m$ with the respective harmonic oscillator Hamiltonian alike Eq. \eqref{eq:Hamiltonian_general_bath}. These modes are coupled to the electronic transition according to Eq. \eqref{eq:HSB1}. The strength of the bilinear coupling of this system (excitons plus intramolecular vibrations) to the phonon bath is described by the coupling constants  $c_{\xi,i}^{(m)}$. In the Ohmic dissipation limit the spectral density for the BO model in case of a single mode with frequency $\omega_{\rm BO}$, reorganization energy $\lambda_{\rm BO}$, and damping $\gamma_{\rm BO}$ is given by~\cite{Mukamel95}
\begin{equation} \label{eq:Brownian_oscillator_spectral_density}
J_{\rm BO}(\omega)=2 \lambda_{\rm BO}  \frac{\omega \omega_{\rm BO}^2\gamma_{\rm BO}}{(\omega_{\rm BO}^2-\omega^2)^2+\gamma_{\rm BO}^2 \omega^2} \,.
\end{equation}

\subsection{Hierarchical Equations of Motion}

In the case of a treatment with HEOM, not only thermal bath modes, but also intramolecular vibrational modes enter as bath components. All bath components need to be  decomposed according to the Matsubara scheme, where coefficients $c_k$ and frequencies $\gamma_k$ enter in the expansion of the correlation function into the series of Matsubara terms
\begin{equation} \label{eq:Matsubara_decomposition_correlation_function}
C(t)=\sum_{k} c_k \exp(-\gamma_k t).
\end{equation}
The term ``Matsubara frequencies'' is commonly used for temperature-dependent frequencies $\gamma_k=2 \pi k/\beta$ with $\beta=1/k_{\rm B}T$, which appear in the Matsubara decomposition of damped bath components and stem from poles of the $\coth$-function accounting for the fluctuation-dissipation relation. In addition so-called explicit terms associated with poles of the spectral density and thus with different definition of the Matsubara decomposition frequencies enter in the series expansion.

The derivation of HEOM has been described in detail in several publications, see e.g. \cite{tanimura06_082001,xu07_031107,schroter15_1,SeMa18_CP_129}.
Therefore, we introduce only the standard terms on the right hand side of the equations of motion for the so-called ``auxiliary density operators'' (ADOs) here.
The ADOs are identified by a subscript set of Matsubara indices, $\boldsymbol{n}$. In the time evolution adjacent ADOs with a difference of one in a single digit of their index patterns are connected to each other.
Each bath component leads to a separate segment of Matsubara indices in the index pattern \cite{LiZhBa14_JCP_134106}.
A formulation of HEOM with the rescaling introduced in Ref. \cite{ShChNa09_JCP_164518} results in
\begin{equation} \label{eq:evaluation_HEOM_scheme_dimer_rescaled}
\begin{split}
&\frac{\partial}{\partial t} \hat{\rho}_{\boldsymbol{n}}=-\left( i {\cal L}_{\rm S} + \sum_{l} \sum_{k} n_{lk} \gamma_k \right) \hat{\rho}_{\boldsymbol{n}} \\
&-i \sum_l \sum_k \sqrt{(n_{lk}+1) |c_k|} \left[ \hat{B}_l^{\dagger} \hat{B}_l, \hat{\rho}_{\boldsymbol{n}^{+}_{lk}} \right] \\
&-i \sum_l \sum_k \sqrt{\frac{n_{lk}}{|c_k|}} \left( c_k \hat{B}_l^{\dagger} \hat{B}_l \hat{\rho}_{\boldsymbol{n}^{-}_{lk}}
-\hat{\rho}_{\boldsymbol{n}^{-}_{lk}} \tilde{c}_k \hat{B}_l^{\dagger} \hat{B}_l \right).
\end{split}
\end{equation}
with ${\cal L}_{\rm S} \hat{\rho}_{\boldsymbol{n}}=\left[ H_{\rm S}, \hat{\rho}_{\boldsymbol{n}} \right]$.
Even though Eq.~(\ref{eq:evaluation_HEOM_scheme_dimer_rescaled}) is formulated in the localized basis, which we rely on in the present work, it can be expressed in the exciton basis as well by applying appropriate transformations to ${\cal L}_{\rm S}$, $\hat{B}_l^{\dagger} \hat{B}_l$ and to the ADOs.
Below, the numerical solution of Eq. \eqref{eq:evaluation_HEOM_scheme_dimer_rescaled} will be used as a reference. It is important to note that the derivation of Eq. \eqref{eq:evaluation_HEOM_scheme_dimer_rescaled} assumes the bath being equilibrated with respect to the electronic ground state

\subsection{Exact Transfer Rates From HEOM}
In order to prepare for the derivation of second-order rates, we briefly summarize the extraction of exact transfer rates according to Ref.~\cite{ZhYa16_JPCA_3241} using  Nakajima-Zwanzig equation.
The Liouville equation with  Liouville space operator $\lsop{L}$ and Liouville space vector $\hat{\boldsymbol{\rho}}$ reads%
\begin{equation} \label{eq:Liouville_equation_general}
\dot{\hat{\boldsymbol{\rho}}}=-i \lsop{L} \hat{\boldsymbol{\rho}}\, .
\end{equation}
Defining Liouville space projectors such that $\lsop{P}+\lsop{Q}={\bf 1}$ yields the Nakajima-Zwanzig equation~\cite{fick90_, May11} 
\begin{equation} \label{eq:QME_general_form}
\begin{split}
&\frac{\partial}{\partial t} \lsop{P} \hat{\boldsymbol{\rho}}(t)=\int_0^t dt' \lsop{K}^{(\rm exact)}(t-t') \lsop{P} \hat{\boldsymbol{\rho}}(t')
-i \lsop{P} \lsop{L} \lsop{P} \hat{\boldsymbol{\rho}}(t) \\
&-i \lsop{P} \lsop{L} \exp(-i \lsop{Q} \lsop{L} t) \lsop{Q} \hat{\boldsymbol{\rho}}(0),
\end{split}
\end{equation}
with
\begin{equation} \label{eq:QME_homogeneous_term}
\lsop{K}^{(\rm exact)}(t)=\lsop{P} \lsop{L} \exp(-i \lsop{Q} \lsop{L} t)
\lsop{Q} \lsop{L} \lsop{P}\,.
\end{equation}

In the context of HEOM $\hat{\boldsymbol{\rho}}$ denotes a Liouville space vector of ADOs and $\lsop{L}$ is a so-called HEOM-space dynamical generator,
which is matrix-valued and contains the Liouville space operators appearing on the right hand side of all hierarchical equations. 
Accordingly, the projection operators in HEOM space, i.e.\ the space of the ADOs, are defined as
\begin{equation} \label{eq:standard_projectors_HEOM_space}
\begin{split}
\lsop{P} \hat{\boldsymbol{\rho}}(t)&=\{ {\mathcal{P}} \hat{\rho}_{\boldsymbol{0}}(t), 0, 0, ... \} \\
\lsop{Q} \hat{\boldsymbol{\rho}}(t)&=\{ {\mathcal{Q}} \hat{\rho}_{\boldsymbol{0}}(t), \hat{\rho}_{\boldsymbol{n} \neq \boldsymbol{0}} \}.
\end{split}
\end{equation}
Application of $\lsop{P}$ leads to selection of a single ADO, namely the system density matrix (equivalent to the reduced density matrix) $\hat{\rho}_{\boldsymbol{0}}$, from which by application of projector ${\mathcal{P}}$
the diagonal elements are selected. The term $-i \lsop{P} \lsop{L} \lsop{P} \hat{\boldsymbol{\rho}}(t)$ is equal to zero~\cite{ZhYa16_JPCA_3241}. 

\subsection{Second-Order Transfer Rates From HEOM} \label{sec:extraction_second-order_rates}
To extract rates from the numerically exact HEOM approach for a treatment of the dissipative dynamics by solving a QME  at a level alike  F\"{o}rster or Redfield theory
(see \cite{YaFl02_CP_355,SeMa17_JCP_174109}), a separation of terms in the hierarchical equations associated with reference and interaction Hamiltonian is required. In the following we will focus on a description in terms of local states (F\"orster case). 

Choosing a perturbative description in terms of local states implies that the off-diagonal part of the system Hamiltonian in the localized basis must be sufficiently small to serve as the interaction Hamiltonian, i.e.\ we have
\begin{align} \label{eq:interaction_Hamiltonian}
\hat{H}'&= \sum_{n \neq m} J_{mn} \hat{B}^{\dagger}_m \hat{B}_n \,.
\end{align}
The reference Hamiltonian contains the diagonal contribution of the system Hamiltonian together with the contributions of bath and system-bath coupling, i.e.
\begin{align}
\hat H_0 &=  \hat{H}_{\rm S,diag} +\hat H_{\rm B} + \hat H_{\rm SB}\,.
\end{align}

Likewise, in the exciton basis second-order rate expressions can be formulated, where the interaction Hamiltonian either contains all system-bath coupling contributions or only the diagonal ones,
depending on whether the assumptions from standard or modified Redfield theory are adopted~\cite{May11}.

HEOM propagation with the reference Hamiltonian is performed according to
\begin{equation} \label{eq:evaluation_HEOM_scheme_dimer_rescaled-relevant}
\begin{split}
&\frac{\partial}{\partial t} \hat{\rho}_{\boldsymbol{n}}=
-i  {\cal L}_{0} \hat{\rho}_{\boldsymbol{n}}=-i \left[ \hat{H}_{\rm S,diag}, \hat{\rho}_{\boldsymbol{n}} \right]-\sum_{l} \sum_{k} n_{lk} \gamma_k \hat{\rho}_{\boldsymbol{n}} \\
&-i \sum_l \sum_k \sqrt{(n_{lk}+1) c_k} \left[ \hat{B}_l^{\dagger} \hat{B}_l, \hat{\rho}_{\boldsymbol{n}^{+}_{lk}} \right] \\
&-i \sum_l \sum_k \sqrt{\frac{n_{lk}}{c_k}} \left( c_k \hat{B}_l^{\dagger} \hat{B}_l \hat{\rho}_{\boldsymbol{n}^{-}_{lk}}
-\hat{\rho}_{\boldsymbol{n}^{-}_{lk}} \tilde{c}_k \hat{B}_l^{\dagger} \hat{B}_l \right).
\end{split}
\end{equation}
The formal solution of this equation can be written using the HEOM-space dynamical generator $\lsop{L}_0$ as follows
\begin{align}
\lop{\rho}(t) = \exp(-i\lsop{L}_0 t)	\lop{\rho}(0)
\end{align}

As the interaction Hamiltonian does not contain any system-bath coupling component, the application of its equivalent Liouville operator 
at the right hand side of the hierarchical equations leads to an expression involving only the same ADO as on the left hand side for each component: 
$\hat{\cal L}' \hat{\rho}_{\boldsymbol{n}}= \left[ \hat{H}', \hat{\rho}_{\boldsymbol{n}} \right]$.

Following Refs. \cite{May11,VaAbMa13} the  QME analogous to Eq.~(\ref{eq:QME_general_form}) in the interaction picture can be split into homogeneous and inhomogeneous parts 
\begin{equation}
\label{eq:QME_hom_inhom}
\frac{\partial}{\partial t} \lsop{P} \hat{\boldsymbol{\rho}}_{\rm I}(t) = \frac{\partial}{\partial t} \lsop{P} \hat{\boldsymbol{\rho}}_{\rm I}(t) \Big|_{\rm hom} +\frac{\partial}{\partial t} \lsop{P} \hat{\boldsymbol{\rho}}_{\rm I}(t) \Big|_{\rm inh} \,.
\end{equation}
For the homogeneous part one obtains
\begin{equation} \label{eq:QME_interaction_picture_tauPrime}
\frac{\partial}{\partial t} \lsop{P} \hat{\boldsymbol{\rho}}_{\rm I}(t) \Big|_{\rm hom}
=-\int_0^t dt' \lsop{K}(t,t') \lsop{P} \hat{\boldsymbol{\rho}}_{\rm I}(t').
\end{equation}
Here, we defined the second-order rate kernel in interaction representation as
\begin{equation}
\label{eq:K2int}
\lsop{K}(t,t') = \lsop{P} \lsop{L}'_{\rm I}(t) 
\lsop{Q} \lsop{L}'_{\rm I}(t') \lsop{P} \,.
\end{equation}
The interaction representation of the Liouville space vector of ADOs and the interaction Liouvillian
are given as $\hat{\boldsymbol{\rho}}_{\rm I}(t)=\exp(i \lsop{L}_0 t) \hat{\boldsymbol{\rho}}(t)$  and $\lsop{L}'_{\rm I}(t)=\exp(i \lsop{L}_0 t) \lsop{L}' \exp(-i \lsop{L}_0 t)$, respectively.

Likewise the inhomogeneous contribution in the interaction picture reads 
\begin{equation} \label{eq:QME_inhomogeneous_term_interaction_picture}
\frac{\partial}{\partial t} \lsop{P} \hat{\boldsymbol{\rho}}_{\rm I}(t) \Big|_{\rm inh}
=-i \lsop{P} \lsop{L}'_{\rm I}(t) \lsop{Q}  \hat{\boldsymbol{\rho}}_{\rm I}(0).
\end{equation}
%


In Eq. \eqref{eq:QME_interaction_picture_tauPrime}
the Markov approximation can be applied by replacing  $\hat{\boldsymbol{\rho}}_{\rm I}(t')\approx \hat{\boldsymbol{\rho}}_{\rm I}(t)$. 

Equation \eqref{eq:K2int}  describes time-dependent second-order rate kernel whose matrix elements can be obtained from successive HEOM propagations according to Eq. \eqref{eq:evaluation_HEOM_scheme_dimer_rescaled-relevant}. As noted above, in the derivation of HEOM the bath is assumed to be in equilibrium with respect to the electronic ground state. For the transfer process this implies an initial condition corresponding to a vertical excitation. For the vibrational degrees of freedom this corresponds to a nonequilibrium situation.  This, however, is in contrast to the assumptions of F\"orster theory. To accommodate the F\"orster limit one has  to account for situations where the bath is equilibrated with respect to the initially excited electronic state~\cite{May11}. In the frame of HEOM this can be accomplished using a polaron transformation (details of the present implementation will be published elsewhere~\cite{seibt20}). 

The polaron transformation shifts the equilibrium position of the oscillator coordinate, e.g.  $x_i$, to a new position $x_i^{(m)}$, which is adjusted to compensate the displacement of a selected oscillator mode coupled to electronic excitation of state $m$~\cite{ReSi96_JCP_1506}. It is introduced by a shift operator formulated as $\hat{D}=\hat{D}(\{x_i^{(m)}\}) =\exp(i \sum_i \hat{p}_i x_i^{(m)} \hat{B}_m^{\dagger} \hat{B}_m)$, where $\hat{p}_i$ is the respective momentum operator. Application to a Hilbert space operator $\hat O$ yields $\hat{\bar O}= \hat{D}^\dagger \hat O \hat D  = {\cal D}^\dagger \hat O$, where the last equality defines the respective Liouville operator. Generalizations to the HEOM space vector of ADOs and HEOM space dynamical generators will be denoted as $\lopBar{\rho}=\lsop{D}^{\dagger}\lop{\rho}$ and  $\lsopBar{L}_0= \lsop{D}^{\dagger}\lsop{L}_0\lsop{D}$, respectively.

A reformulation of Eq.~(\ref{eq:QME_hom_inhom}) using the  polaron transformation leads to
\begin{equation} \label{eq:QME_homogeneous_term_interaction_picture_with_pol_trans}
\begin{split}
&\frac{\partial}{\partial t} \lsopBar{P}\lsop{D}^{\dagger} \hat{\boldsymbol{\rho}}_{\rm I}(t) \Big|_{\rm hom}= \frac{\partial}{\partial t} \lsopBar{P}\lopBar{\rho}_{\rm I}(t) \Big|_{\rm hom}\\
&=-\int_0^t dt' \lsopBar{K}(t,t') \lsopBar{P}\lopBar{\rho}_{\rm I}(t') \\
\end{split}
\end{equation}
with the second-order rate kernel
\begin{align} \label{eq:rate_kernel_with_pol_trans}
\lsopBar{K}(t,t') &= 	\lsopBar{P}\lsop{D}^{\dagger} 
\lsop{L}'_{\rm I}(t) \lsop{D} \lsopBar{Q}
\lsop{D}^{\dagger} \lsop{L}'_{\rm I}(t') \lsop{D} \lsopBar{P}
\end{align}

For the inhomogeneous term one obtains
\begin{equation} \label{eq:QME_inhomogeneous_term_interaction_picture_with_pol_trans}
\frac{\partial}{\partial t} \lsopBar{P}\lopBar{\rho}_{\rm I}(t) \Big|_{\rm inh}
=-i \lsopBar{P}\lsop{D}^{\dagger} \lsop{L}'_{\rm I}(t)
\lsop{D} \lsopBar{Q} \lopBar{\rho}_{\rm I}(0).
\end{equation}
Note that the action of the polaron-transformed projector $\lsopBar{P}$ onto the transformed ADO vector $\lopBar{\rho}_{\rm I}$ is alike Eq. \eqref{eq:standard_projectors_HEOM_space}. 
Using this definition it follows that $\lsop{L}'_{\rm I}(t') \lsop{D} \lsopBar{P}\lopBar{\rho}_{\rm I}(t')$ gives a non-diagonal reduced density matrix such that $ \lsop{D} \lsop{Q}\lsop{D}^{\dagger}=  \lsop{D} 
\lsop{D}^{\dagger}={\bf 1}$.

Again, the Markov approximation can be applied by setting $\lopBar{\rho}_{\rm I}(t')=\lopBar{\rho}_{\rm I}(t)$.

%
\subsection{Second-Order Transfer Rates From Cumulant Expansion}
\label{sec:cumulant}

The generic expressions for rate kernels in Liouville space can also be evaluated via the cumulant expansion  in second-order approximation  to obtain a formulation in terms of line shape functions \cite{YaFl02_CP_355,SeMa17_JCP_174109}. 
The approach of using cumulant expansion for rate calculation turns out to yield equivalent results as the calculation of the rates with HEOM, at least for our model and for the ways of calculating second-order transfer rates we consider in this work. However, there are cases where it is not practicable to formulate rate expressions via cumulant expansion in terms of analytical expressions, so that the approach of rate calculation with HEOM is preferable, as will be discussed in Sec. \ref{sec:discussion}. In this work we concentrate on the situation where both approaches can be applied, as a basis for further investigations.
Furthermore, as the rate expressions in terms of line shape functions are more comprehensible than the generic Liouville space expressions,
they are useful for an interpretation of differences arising from specific aspects in the calculation of transfer rates. 

The QME for the reduced density matrix, which is obtained by application of the projection operator ${\cal P}$ to the full density matrix $\hat{\rho}$ as
\begin{align}
{\cal P} \hat{\rho}=\sum_m \hat{B}^{\dagger}_m \hat{B}_m  {\rm Tr}_{\rm B} \{\langle m| \hat{\rho} | m \rangle  \} \hat{W}_{\rm B}	
\end{align}
with $\hat{W}_{\rm B}$ being the statistical operator of the bath, is formulated in the interaction picture. 
One obtains
\begin{align}
\label{eq:QME_system}
\frac{\partial}{\partial t} {\cal P} \hat{\rho}_{\rm I}(t)& = \frac{\partial}{\partial t} {\cal P} \hat{\rho}_{\rm I}(t) \Big|_{\rm hom} +\frac{\partial}{\partial t} {\cal P} \hat{\rho}_{\rm I}(t) \Big|_{\rm inh}	
\end{align}

where the homogeneous and inhomogeneous contribution are given as
\begin{align}
\label{eq:QME_R-term}
\frac{\partial}{\partial t} {\cal P} \hat{\rho}_{\rm I}(t) \Big|_{\rm hom} &= 	-\int_0^t dt' {\cal K}^{\rm (c)}(t,t') {\cal P} \hat{\rho}_{\rm I}(t')
\end{align}
and 
\begin{align}
\label{eq:QME_I-term}
\frac{\partial}{\partial t} {\cal P} \hat{\rho}_{\rm I}(t) \Big|_{\rm inh} = 	-i \sum_{m \neq n} | m \rangle \langle n | {\rm Tr}_{\rm B}  \{ [{\cal L}'_{\rm I}(t)]_{mn} \} \hat{W}_{\rm B} {\cal Q} \hat{\rho}_{\rm I}(0),
\end{align}
respectively.
Here, the second-order rate kernel is 
\begin{equation} 
\begin{split}
&{\cal K}^{\rm (c)}(t,t')=\sum_m \hat{B}^{\dagger}_m \hat{B}_m {\rm Tr}_{\rm B}  \{ [{\cal L}'_{\rm I}(t){\cal Q} {\cal L}'_{\rm I}(t')]_{mm} \} \hat{W}_{\rm B} \\
&=\sum_m \hat{B}^{\dagger}_m \hat{B}_m 
{\rm Tr}_{\rm B} \{ [\exp(i {\cal L}_0 t) \hat{\cal L}' \exp(-i {\cal L}_0 t) \\
&\times {\cal Q} \exp(i {\cal L}_0 t') {\cal L}' \exp(-i {\cal L}_0 t')]_{mm} \} \hat{W}_{\rm B} \,.
\end{split}
\end{equation}
The rate expression from the homogeneous contribution given in Eq.~(\ref{eq:QME_R-term}) is equivalent to the one from Eq.~(\ref{eq:K2int}) in HEOM space. 
Note that Eq.~(\ref{eq:QME_system}) is formulated as a non-Markovian QME, but due to representation in the interaction picture it can be turned into a Markovian one by simply replacing the time argument $t'$ of the reduced density matrix with $t$, so that the reduced density matrix can be drawn out of the integral.
Depending on the assumption about initial equilibration, $\hat{W}_{\rm B}$ corresponds to the bath equilibrium density matrix in the ground- or the electronically excited initial state of the transfer process.

In a F\"{o}rster-type description of population transfer in the localized basis, the off-diagonal part of the system Hamiltonian composed of excitonic couplings is identified with the interaction Hamiltonian, which accordingly consists of non-zero off-diagonal contributions
\begin{equation}
\hat{H}'_{mn}=J_{m n} \hat{B}^{\dagger}_m \hat{B}_n ,
\end{equation}
equivalent to the sum components of the expression given in Eq.~(\ref{eq:interaction_Hamiltonian}). The different diagonal components $\hat{H}_{0,mm}$ of the reference Hamiltonian contain
the vertical electronic excitation energies $\epsilon_m+\lambda_m$, the bath Hamiltonian from Eq.~(\ref{eq:Hamiltonian_general_bath}) and the gap coordinate operator $\hat u_m$ from the system-bath coupling Hamiltonian given in Eq.~(\ref{eq:HSB1}).

First the case of initial equilibration in the electronic ground state with $\hat{W}_{\rm B}=\hat{W}_{{\rm B},g}=\prod_{m} \hat{W}_{{\rm B},g_m}$ 
will be discussed. The corresponding rate expression has already been derived in Ref.~\cite{SeMa17_JCP_174109}. We repeat the derivation here to draw connections to additional aspects, which we will discuss later. In the considered case with a \emph{non-equilibriated bath} after electronic excitation from the ground state, in Hilbert-space formulation the rates for transfer from monomer $m$ to monomer $n$, which correspond to tensor elements in Eq.~(\ref{eq:QME_R-term}), can be identified as
\begin{equation} \label{eq:Foerster_rate_homogeneous_term_matrix_elements_general_expression_noneq_no_int_pic}
\begin{split}
&{\cal K}^{\rm (c, noneq)}_{nn,mm}(t,t')=2 \Re \bigg( {\rm Tr}_{\rm B} \{ \exp(i \hat{H}_{0,m m} t) \hat{H}'_{m n} \\
&\times  \exp(-i \hat{H}_{0,n n} t) \exp(i \hat{H}_{0,n n} t') \hat{H}'_{n m} \\
&\times \exp(-i \hat{H}_{0,m m} t') \} \hat{W}_{{\rm B},g} \bigg).
\end{split}
\end{equation}
This expression can be evaluated by using the cumulant expansion technique, where the correlation function of the time-dependent gap coordinates 
\begin{equation} \label{eq:time_evolution_energy_gap_coordinates}
\hat{u}_{{\rm I},m}(\tau)=\exp(i \hat{H}_{\rm B} \tau) \hat{u}_{m} \exp(-i \hat{H}_{\rm B} \tau).
\end{equation}
enters in the line shape functions
\begin{equation} \label{eq:line_shape_function_correlation_function}
\begin{split}
&g_{m}(\tau)=\int_{0}^{\tau} d\tau' \int_{0}^{\tau'} d\tau'' {\rm Tr}_{\rm B} \{ \hat{u}_{I,m}(\tau'') \hat{u}_{I,m}(0) \} \\
&=\int_{0}^{\tau} d\tau' \int_{0}^{\tau'} d\tau'' C_m(\tau'').
\end{split}
\end{equation}
The respective rate expression in terms of line shape functions results as 
\begin{equation} \label{eq:Foerster_rate_homogeneous_term_line_shape_functions_noneq}
\begin{split}
&{\cal K}^{\rm (c, noneq)}_{nn,mm}(t,t')= 2 |J_{mn}|^2 \Re \bigg( \exp(i (\epsilon_m-\epsilon_n) (t-t')) \\
&\times \exp(i (\lambda_m-\lambda_n) (t-t')) \exp \left[ -g_{n}(t-t')-g_{m}(t-t') \right. \\
&\left. +2i \Im(g_m(t))-2i \Im(g_m(t')) \right] \bigg).
\end{split}
\end{equation}
The inhomogeneous term is obtained as
\begin{align}
\label{eq:QME_I-term}
\frac{\partial}{\partial t} {\cal P} \hat{\rho}_{\rm I}(t) \Big|_{{\rm inh},nn,mm}^{\rm (c, noneq)} &= i J_{mn} \exp(-i (\epsilon_m-\epsilon_n) t) \nonumber\\
\times \exp(-i (\lambda_m-\lambda_n) t) 
& \exp \left( -g^{*}_{n}(t)-g_{m}(t) \right).
\end{align}

In the case with thermal equilibration in the excited initial state of the transfer process the density matrix of the bath is identified with 
$\hat{W}_{\rm B}=\hat{W}_{{\rm B},m}=\hat{W}_{{\rm B},e_m} \prod_{n \neq m} \hat{W}_{{\rm B},g_n}$.
In the context of the cumulant expansion technique, instead of applying the polaron transformation according to Eq.~(\ref{eq:rate_kernel_with_pol_trans})
we adopt the treatment of initial equilibration proposed in \cite{ZhMeCh98_JCP_7763}. In this way one can express initial thermal equilibration via
\begin{equation} \label{eq:initial_equilibration_excited_state}
\hat{W}_{{\rm B},m}=\lim_{\tau \to \infty} \exp(-i {\mathcal{L}}_0 \tau) \hat{W}_{{\rm B},g} \hat{B}_m^{\dagger} \hat{B}_m.
\end{equation}
Insertion into the rate expression yields
\begin{equation} \label{eq:Foerster_rate_homogeneous_term_matrix_elements_general_expression_std}
\begin{split}
&{\cal K}^{\rm (c,equi)}_{nn,mm}(t,t')=2 \Re \bigg( {\rm Tr}_{\rm B} \{ \exp(i \hat{H}_{0,m m} t) \hat{H}'_{m n} \\
&\times \exp(-i \hat{H}_{0,n n} t) \exp(i \hat{H}_{0,n n} t') \hat{H}'_{n m} \\
&\times \exp(-i \hat{H}_{0,m m} t') \} \hat{W}_{{\rm B},m} \bigg) \\
&=\lim_{\tau \to \infty} 2 \Re \bigg( {\rm Tr}_{\rm B} \{ \exp(-i \hat{H}_{0,m m} \tau) \exp(i \hat{H}_{0,m m} t) \\
&\times \hat{H}'_{m n} \exp(-i \hat{H}_{0,n n} t) \exp(i \hat{H}_{0,n n} t') \hat{H}'_{n m} \\
&\times \exp(-i \hat{H}_{0,m m} t') \exp(i \hat{H}_{0,m m} \tau) \} \hat{W}_{{\rm B},g} \bigg).
\end{split}
\end{equation}
By applying the cumulant expansion technique and by identifying the derivative of the imaginary part of the line shape functions in the limit of large time arguments with the negative reorganization energy $\lambda_m$ \cite{ZhMeCh98_JCP_7763}, one obtains the rate for the \textit{equilibrated bath}, i.e. F\"orster case
\begin{equation} \label{eq:Foerster_rate_homogeneous_term_line_shape_functions_std}
\begin{split}
&{\cal K}^{\rm (c,equi)}_{nn,mm}(t,t')= 2 |J_{mn}|^2 \Re \bigg( \exp(i (\epsilon_m-\epsilon_n) (t-t')) \\
&\exp(i (-\lambda_m-\lambda_n) (t-t')) \exp \left( -g_{n}(t-t')-g_{m}(t-t') \right) \bigg).
\end{split}
\end{equation}

The inhomogeneous term is obtained as
\begin{align}
\label{eq:QME_I-term}
&\frac{\partial}{\partial t} {\cal P} \hat{\rho}_{\rm I}(t) \Big|_{{\rm inh},nn,mm}^{\rm (c,equi)} = -i J_{mn} \exp(-i (\epsilon_m-\epsilon_n) t)\nonumber\\
&\times  \exp(i (\lambda_m-\lambda_n) t) 
 \exp \left( -g_{n}(t)-g^{*}_{m}(t) \right).
\end{align}

It is interesting to note that Eq. \eqref{eq:Foerster_rate_homogeneous_term_matrix_elements_general_expression_std} is equivalent to the tensor elements in the HEOM-space formulation from Eq.~(\ref{eq:rate_kernel_with_pol_trans}) where the polaron transformations has been applied.
As the displacement in the initially excited state $m$ is compensated by the polaron transformation or, equivalently, by applying Eq.~(\ref{eq:initial_equilibration_excited_state})
in the context of the cumulant expansion technique, the respective bath oscillators assigned to excitation of monomer $m$ are expressed as unshifted oscillators.
Therefore, the reference Liouville operators and related time evolution operators, which appear in a back-transformation from the interaction to the Schr\"odinger picture, are commutable with the projectors
and can thus be drawn into the trace expression. After such a back-transformation the Liouville-space formulation of the rate expression in the Schr\"odinger picture becomes
\begin{equation} \label{eq:QME_R-term_outside_interaction_picture}
\begin{split}
&{\cal K}^{\rm (c,S)}(t,t')=\sum_m \hat{B}^{\dagger}_m \hat{B}_m
{\rm Tr}_{\rm B} \{ [\exp(-i {\cal L}_0 t) \exp(i {\cal L}_0 t) {\cal L}'  \\
&\times \exp(-i {\cal L}_0 t) \hat{\cal Q} \exp(i {\cal L}_0 t') {\cal L}'  \exp(-i {\cal L}_0 t') \exp(i {\cal L}_0 t)]_{mm} \} \\
&\times \hat{W}_{{\rm B},m} \\
&=\sum_m \hat{B}^{\dagger}_m \hat{B}_m
{\rm Tr}_{\rm B} \{ [ {\cal L}' \exp(-i {\cal L}_0 t) {\cal Q} \exp(i \hat{\cal L}_0 t') {\cal L}' \\
&\times \exp(-i {\cal L}_0 t') \exp(i {\cal L}_0 t)]_{mm} \} \hat{W}_{{\rm B},m}\,.
\end{split}
\end{equation}
In Hilbert-space one obtains
\begin{equation} \label{eq:Foerster_rate_homogeneous_term_matrix_elements_general_expression_noneq_no_int_pic}
\begin{split}
&{\cal K}^{\rm (c,equi,S)}_{nn,mm}(t,t')=2 \Re \bigg( {\rm Tr}_{\rm B} \{ \hat{H}'_{m n} \exp(-i \hat{H}_{0,n n} t) \\
&\times \exp(i \hat{H}_{0,n n} t') \hat{H}'_{n m} \exp(-i \hat{H}_{0,m m} t') \exp(i \hat{H}_{0,m m} t) \} \\
&\times \hat{W}_{{\rm B},m} \bigg) \\
&=\lim_{\tau \to \infty} 2 \Re \bigg( {\rm Tr}_{\rm B} \{ \exp(-i \hat{H}_{0,m m} \tau) \hat{H}'_{m n} \exp(-i \hat{H}_{0,n n} t) \\
&\times \exp(i \hat{H}_{0,n n} t') \hat{H}'_{n m} \exp(-i \hat{H}_{0,m m} t')  \exp(i \hat{H}_{0,m m} t) \} \\
&\times \exp(i \hat{H}_{0,m m} \tau) \hat{W}_{{\rm B},g}  \bigg).
\end{split}
\end{equation}
Evaluation by using the cumulant expansion technique yields Eq.~(\ref{eq:Foerster_rate_homogeneous_term_line_shape_functions_std}), i.e.\ the same expression as in the interaction picture.
However, this finding is only valid under the assumption of thermal equilibration in the excited initial state of the transfer process. Otherwise the operators appearing in the back-transformation
from the interaction picture cannot be taken as a part of the trace expression. If despite the non-commutability with the projectors the time-evolution operators from the back-transformation are included into the non-equilibrium case, it turns out that the resulting rate expression in terms of line shape functions differs from the one given in Eq.~(\ref{eq:Foerster_rate_homogeneous_term_line_shape_functions_noneq}) by the additional line shape function contributions
$2i \Im(g_m(t-t'))-2i \Im(g_m(t))+2i \Im(g_m(t'))$ in the argument of the exponential. Such terms can be attributed to vibrational relaxation of the bath \cite{SePu14_JCP_114106}. Disregarding them implies the assumption of thermal equilibration with respect to the electronic ground state during the whole transfer process -- an assumption which reminds of the standard Redfield treatment in the exciton basis. In fact, rates calculated in the localized basis under this assumption can yield meaningful results in very specific regimes, i.e.\ where the transfer dynamics is governed by fluctuations without substantial involvement of dissipation in the sense of vibrational relaxation.
Therefore, below this scenario will be called \textit{fluctuation-only case}.
%
\section{Results} \label{sec:results}
%
In the following we will illustrate the approach presented in Section \ref{sec:theoretical_background} for a molecular dimer, putting emphasis on the effect of the various approximations. 
For reference numerically exact HEOM propagations for the full density matrix have been performed. Here, the initial condition corresponds to the vertical excitation of monomer 1 (vibrationally non-equilibrated). 
For the calculation of the population dynamics with HEOM  
we used a time step of $\Delta t=\unit[0.0625]{fs}$.
Besides the explicit terms from the Matsubara decomposition of Debye-Drude and Brownian spectral density, we took a single additional term with the lowest Matsubara frequency $\gamma_1=2 \pi/\beta$ into account, which for a temperature of $\unit[300]{K}$ has a value of $\gamma_1=\unit[1310]{cm^{-1}}$ and is thus considerably larger than the parameters $\omega_{\rm c}=\unit[50]{cm^{-1}}$ and $\omega_{\rm BO}=\unit[200]{cm^{-1}}$ entering in the respective spectral densities.
The truncation order, i.e.\ an upper bound for the values of the Matsubara index digits from the subscript index pattern of the ADOs, which may not be exceeded without the respective ADO being disregarded in the propagation, was set to a value of $\unit[10]{}$.

As the time propagations involved in the calculation of the rates with HEOM make this approach less efficient, we calculated the rates for integration of the non-Markovian QME from the analytical expressions in terms of line shape functions given in Eq. \eqref{eq:Foerster_rate_homogeneous_term_line_shape_functions_noneq} (non-equilibrium case) 
and Eq. \eqref{eq:Foerster_rate_homogeneous_term_line_shape_functions_std} (equilibrium case). 
Further, results for the approximate kernel Eq. \ref{eq:Foerster_rate_homogeneous_term_matrix_elements_general_expression_noneq_no_int_pic} (fluctuation-only case) will be given.

Concerning the system parameters we will in particular vary the ratio between energy offset $\Delta \epsilon=\epsilon_1-\epsilon_2$ and the Coulomb coupling $J_{12}$. Propagation time will be given in units of the inverse excitonic energy gap, i.e. $T=2 \pi/\Omega$  with $\Omega=\sqrt{(\Delta \epsilon)^2 +4 J_{12}^2}$. Instead of showing population dynamics in both excited states, for clarity we display their difference, $P_1(t)-P_2(t)$, which contains information about oscillatory dynamics and asymptotic behavior.
\begin{figure}[th] 
\includegraphics*[width=0.8\columnwidth]{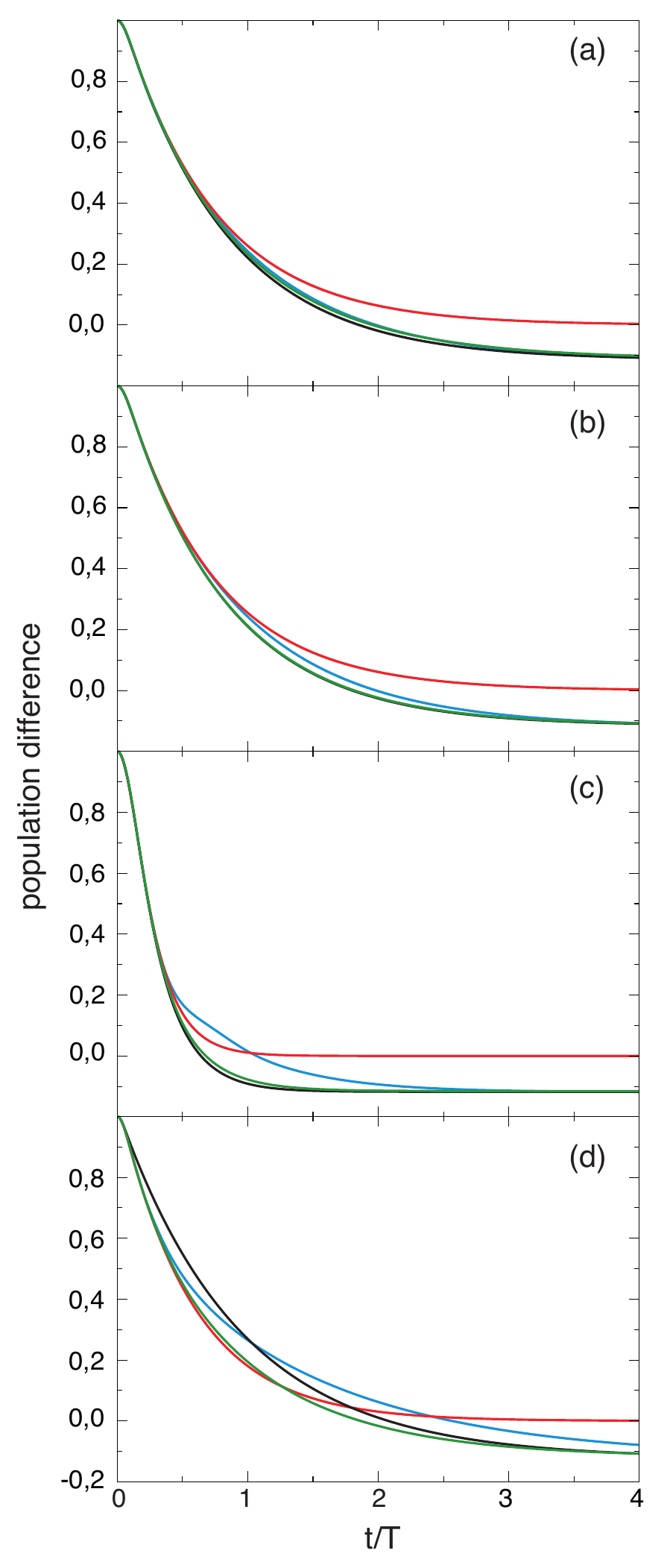}
\caption{Population difference dynamics of the heterodimer model ($\Delta \epsilon=\unit[50]{cm^{-1}}$, $T=\unit[468]{fs}$ (a-b), $T=\unit[296]{fs}$ (c-d)) coupled to a Debye-Drude thermal bath ($\lambda_{\rm DD}=\unit[50]{cm^{-1}}$ (a-c)). The parameters are (a) $\omega_{c}=\unit[12.5]{cm^{-1}}$ and $J_{12}=\unit[25]{cm^{-1}}$, (b)  $\omega_{c}=\unit[50]{cm^{-1}}$ and $J_{12}=\unit[25]{cm^{-1}}$, (c)  $\omega_{c}=\unit[50]{cm^{-1}}$ and $J_{12}=\unit[50]{cm^{-1}}$, (d)  $\omega_{c}=\unit[50]{cm^{-1}}$, $J_{12}=\unit[50]{cm^{-1}}$, and $\lambda_{\rm DD}=\unit[200]{cm^{-1}}$.  Color code: light blue -- exact HEOM, green - non-equilibrium case Eq. \eqref{eq:Foerster_rate_homogeneous_term_line_shape_functions_noneq}, black -- equilibrium case Eq. \eqref{eq:Foerster_rate_homogeneous_term_line_shape_functions_std}, red -- fluctuation-only case Eq. \eqref{eq:Foerster_rate_homogeneous_term_matrix_elements_general_expression_noneq_no_int_pic}.}
\label{fig:1}
\end{figure}

\subsection{Debye-Drude Case}
In Fig. \ref{fig:1} results from exact HEOM propagation are contrasted to non-equilibrium,  Eq. \eqref{eq:Foerster_rate_homogeneous_term_line_shape_functions_noneq},  and equilibrium, Eq. \eqref{eq:Foerster_rate_homogeneous_term_line_shape_functions_std}, second-order rate approaches for the case $\lambda_{\rm DD}=\Delta \epsilon$. In case of weak Coulomb coupling $J_{12}/\lambda_{\rm DD}=1/2$ shown in panels (a) and (b) the exact HEOM dynamic obeys a simple close to exponential decay of the initially excited state. The two panels differ by the choice of the bath correlation time which is $\unit[424]{fs}$ and $\unit[106]{fs}$ in panel (a) and (b), respectively. As a consequence there is an almost perfect agreement between the exact and the non-equilibrium results in panel (a), whereas the population decay is too fast in the equilibrium case. For the shorter correlation time in panel (b) non-equilibrium and equilibrium results essentially agree with each other, but show a decay which is faster than the exact reference. Because the the ratio $J_{12}/\lambda_{\rm DD}$ did not change compared to panel (a) the difference should be attributed to higher-order effects, which in the present case are visible only for the shorter correlation time. Of course, such higher-order effects will become more pronounced with increasing the Coulomb coupling strength. This is shown in Fig. \ref{fig:1}(c) for the case $J_{12}/\lambda_{\rm DD}=1$. As expected, both equilibrium and non-equilibrium rates increase, leading to a fast population decay. In contrast, the decay slows down in the exact case where  a pronounced non-exponential behavior points to partially coherent population exchange between the two monomers.

Figure \ref{fig:1}(d) shows the effect of increasing the reorganization energy such that $J_{12}/\lambda_{\rm DD}=0.25$ while keeping the other parameters as in panel (c). Considering $J_{12}/\lambda_{\rm DD}$ as the small parameter one should be closer to the second-order regime. Indeed the coherent oscillations seen in panel (c) disappear. On the other hand, it takes more time to equilibrate, which leads to an overall slow-down of the population transfer. Although both equilibrium and non-equilibrium results show a similar slow-down, neither of them is in quantitative agreement with the exact HEOM reference.

Figure \ref{fig:1} also contains the results obtained for the fluctuation-only case, Eq. \eqref{eq:Foerster_rate_homogeneous_term_matrix_elements_general_expression_noneq_no_int_pic}. It turns out that while the initial decay is reasonably reproduced by this approximation, the lack of proper description of vibrational relaxation leads to a wrong asymptotic behavior, i.e. equilibration cannot be described with this method.
\subsection{Multimode Brownian Oscillator Case}
The BO model will be discussed for the underdamped limit where $\omega_{\rm BO}>\gamma_{\rm BO}$. 
For the frequency we will take $\omega_{\rm BO}=\unit[200]{cm^{-1}}$ whereas $\gamma_{\rm BO}$ and $S_{\rm BO}$ will be varied.

\begin{figure}[h] 
\includegraphics*[width=0.8\columnwidth]{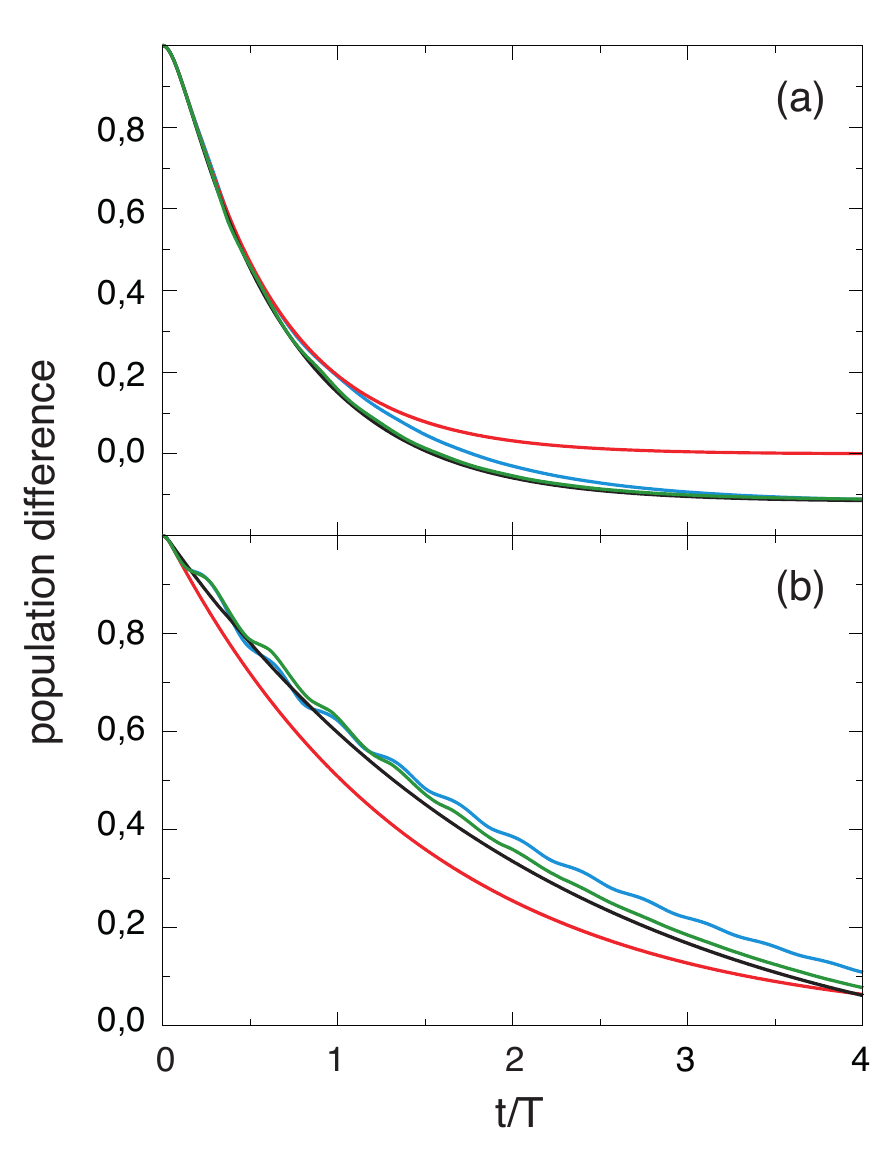}
\caption{Population difference dynamics of the heterodimer model ($\Delta \epsilon=\unit[50]{cm^{-1}}$, $J_{12}=\unit[25]{cm^{-1}}$, $T=\unit[468]{fs}$) coupled to a BO bath ($\omega_{\rm BO}=\unit[200]{cm^{-1}}$)  for (a) $\gamma_{\rm BO}=\unit[50]{cm^{-1}}$, $S_{\rm BO}=0.25$, (b) $\gamma_{\rm BO}=\unit[12.5]{cm^{-1}}$, $S_{\rm BO}=1.0$.  Color code: light blue -- exact HEOM, green -- non-equilibrium case Eq. \eqref{eq:Foerster_rate_homogeneous_term_line_shape_functions_noneq}, black -- equilibrium case Eq. \eqref{eq:Foerster_rate_homogeneous_term_line_shape_functions_std}, red -- fluctuation-only case Eq. \eqref{eq:Foerster_rate_homogeneous_term_matrix_elements_general_expression_noneq_no_int_pic}.}
\label{fig:2}
\end{figure}

In Fig. \ref{fig:2} the population dynamics is shown for the same exciton parameters as in Fig. \ref{fig:1}(a,b). In panel (a) we have chosen $S_{\rm BO}=0.25$ which gives $\Delta \epsilon=\lambda_{\rm BO}$ and $\gamma_{\rm BO}= \unit[50]{cm^{-1}}$. For such a small Huang-Rhys factor and thus small ratio $J_{12}/\omega_{\rm BO}$ the energy spectrum is excitonically dominated (cf. classification in Ref.~\cite{schroter15_1}). Interestingly, the behavior in Fig. \ref{fig:2}(a) pretty much resembles the one in 
Fig. \ref{fig:1}(b). At first glance this is surprising, but one should note that  $\lambda_{\rm DD}= \lambda_{\rm BO}$, and $\omega_{\rm c}=\gamma_{\rm BO}$. 
In order to see effects of the vibrations the Huang-Rhys factor has to be increased. This is shown in Fig. \ref{fig:2}(b) for $S_{\rm BO}=1.0$ where we also decreased the damping to $\gamma_{\rm BO}=\unit[12.5]{cm^{-1}}$. Here, the exact and the non-equilibrium cases show rather similar small amplitude modulation of the population decay. Naturally, this is not seen for the equilibrium case. The fluctuation-only case doesn't give a proper description for both parameter sets. For the large value of $\gamma_{\rm BO}$ the rapid relaxation cannot be reproduced, whereas in case of the small $\gamma_{\rm BO}$ and larger $S_{\rm BO}$ the effect of the fluctuations on the population decay is overestimated. 

\begin{figure}[h] 
\includegraphics*[width=0.8\columnwidth]{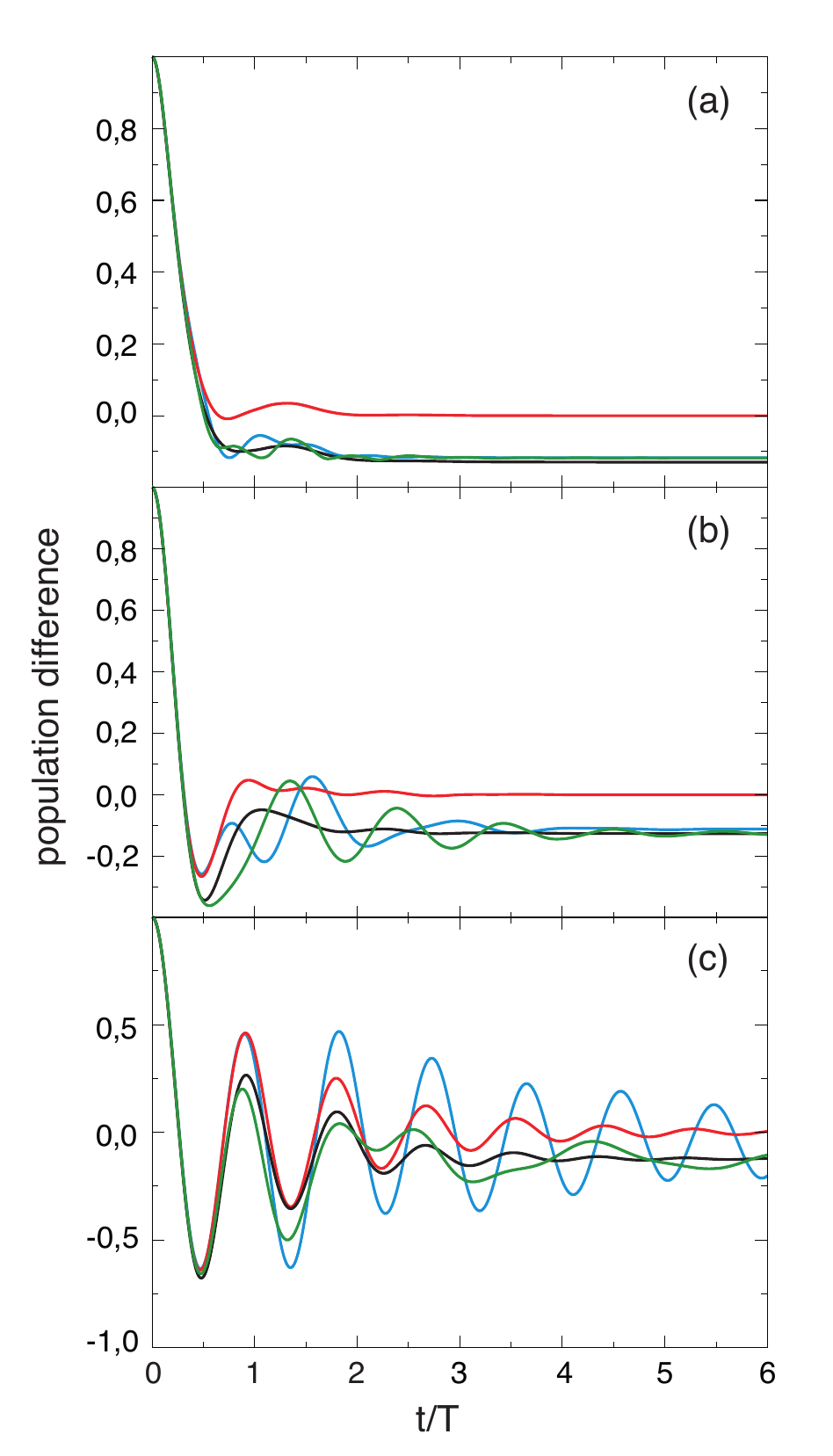}
\caption{Population difference dynamics of the heterodimer model ($\Delta \epsilon=\unit[50]{cm^{-1}}$) coupled to a BO bath ($\omega_{\rm BO}=\unit[200]{cm^{-1}}$, $\gamma_{\rm BO}=\unit[50]{cm^{-1}}$, $S_{\rm BO}=0.25$)  for (a) $J_{12}=\unit[50]{cm^{-1}}$ ($T=\unit[296]{fs}$), (b)  $J_{12}=\unit[100]{cm^{-1}}$ ($T=\unit[161]{fs}$), and (c)  $J_{12}=\unit[200]{cm^{-1}}$ ($T=\unit[82]{fs}$).  Color code: light blue -- exact HEOM, green -- non-equilibrium case Eq. \eqref{eq:Foerster_rate_homogeneous_term_line_shape_functions_noneq}, black -- equilibrium case Eq. \eqref{eq:Foerster_rate_homogeneous_term_line_shape_functions_std}, red -- fluctuation-only case Eq. \eqref{eq:Foerster_rate_homogeneous_term_matrix_elements_general_expression_noneq_no_int_pic}.}
\label{fig:3}
\end{figure}

In the following we will proceed with $S_{\rm BO}=0.25$ ($\lambda_{\rm BO}=\unit[50]{cm^{-1}}$) and $\gamma_{\rm BO}=\unit[50]{cm^{-1}}$ and explore the range  of increasing Coulomb couplings. Fig. \ref{fig:3}(a) shows the case $J_{12}/\lambda_{\rm BO}=1$. The population transfer essentially proceeds upon the first excursion with exact, non-equilibrium, and equilibrium cases giving rather similar results. 
Increasing the coupling to $J_{12}/\lambda_{\rm BO}=2$ in Fig. \ref{fig:3}(b) one observes some population oscillations, whose amplitude and phase and not very well captured by the second-order non-equilibrium and equilibrium approaches. Both, however, reproduce approximately the correct equilibration at longer times. 
The fluctuation-only approach fails in giving the correct equilibration. However, the first oscillation is rather well reproduced in case of the stronger coupling, where transfer dominates the relaxation. One wouldn't expect to get anything meaningful from the second-order approaches for $J_{12}/\lambda_{\rm BO}=4$. Still the results shown Fig. \ref{fig:3}(c) show that the first two oscillation are reasonably reproduced. Interestingly, the fluctuation-only case performs best, for the same reason as in panel (b). 
For longer times all approximate approaches predict a too fast equilibration.

\begin{figure}[h] 
\includegraphics*[width=0.8\columnwidth]{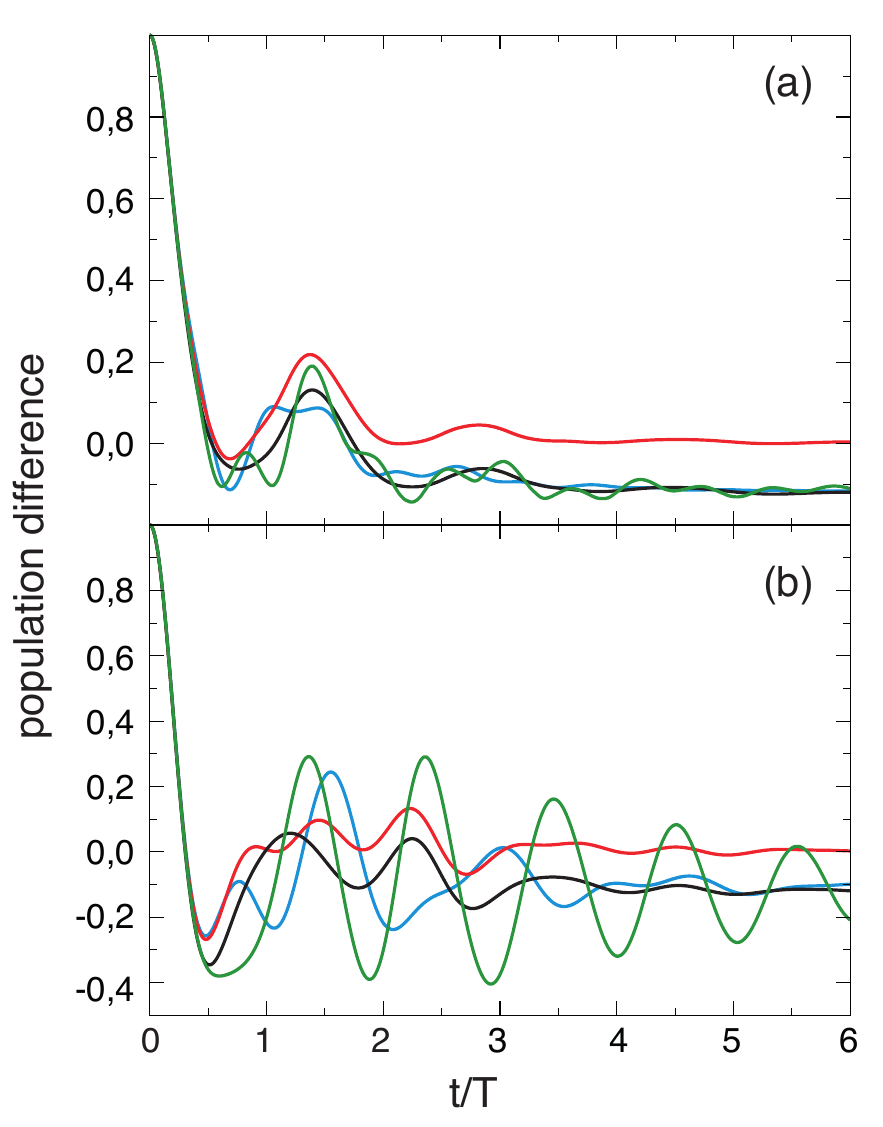}
\caption{Population difference dynamics of the heterodimer model ($\Delta \epsilon=\unit[50]{cm^{-1}}$) coupled to a BO bath ($\omega_{\rm BO}=\unit[200]{cm^{-1}}$, $\gamma_{\rm BO}=\unit[20]{cm^{-1}}$, $S_{\rm BO}=0.25$)  for (a) $J_{12}=\unit[50]{cm^{-1}}$ ($T=\unit[296]{fs}$) and (b)  $J_{12}=\unit[100]{cm^{-1}}$ ($T=\unit[162]{fs}$).  Color code: light blue -- exact HEOM, green -- non-equilibrium case Eq. \eqref{eq:Foerster_rate_homogeneous_term_line_shape_functions_noneq}, black -- equilibrium case Eq. \eqref{eq:Foerster_rate_homogeneous_term_line_shape_functions_std}, red -- fluctuation-only case Eq. \eqref{eq:Foerster_rate_homogeneous_term_matrix_elements_general_expression_noneq_no_int_pic}.}

\label{fig:4}
\end{figure}

In order to explore how the relaxation rate affects the dynamics we show in Fig. \ref{fig:4}(a) and (b) the case $J_{12}/\lambda_{\rm BO}=1$ and $J_{12}/\lambda_{\rm BO}=2$, respectively, for $\gamma_{\rm BO}=\unit[20]{cm^{-1}}$. Compared to Fig. \ref{fig:3}(a) the decrease of the relaxation rate leads to the appearance of about two periods of population oscillation, which are qualitatively  captured by the second-order non-equilibrium and equilibrium rates. The slower relaxation even causes the fluctuation-only description to become reasonable for the first oscillation period. Panel (b) has to be compared with Fig. \ref{fig:3}(b). First, we notice the overall increase of oscillatory behavior. More interesting is the fact that different approximations work well in different time ranges. Initially, again the fluctuation-only approach is in good agreement with the exact reference. Asymptotically, however, it doesn't give the correct equilibration. The equilibrium rate result, on the other hand, gives a reasonable description from around $t=3T$, where the exact solution carries already the signatures of almost completed vibrational relaxation. The non-equilibrium results are generally poor, although they correspond to the proper initial conditions as compared to the exact reference. In the early evolution there is still some similarity with the exact result from HEOM, but from $t=2T$ onwards the decay of the oscillation amplitude is considerably faster in the latter case, where also modulation with an additional frequency component (obviously the BO central frequency) becomes recognizable. This should attributed to higher-order effects.
\begin{figure}[h] 
\includegraphics*[width=0.8\columnwidth]{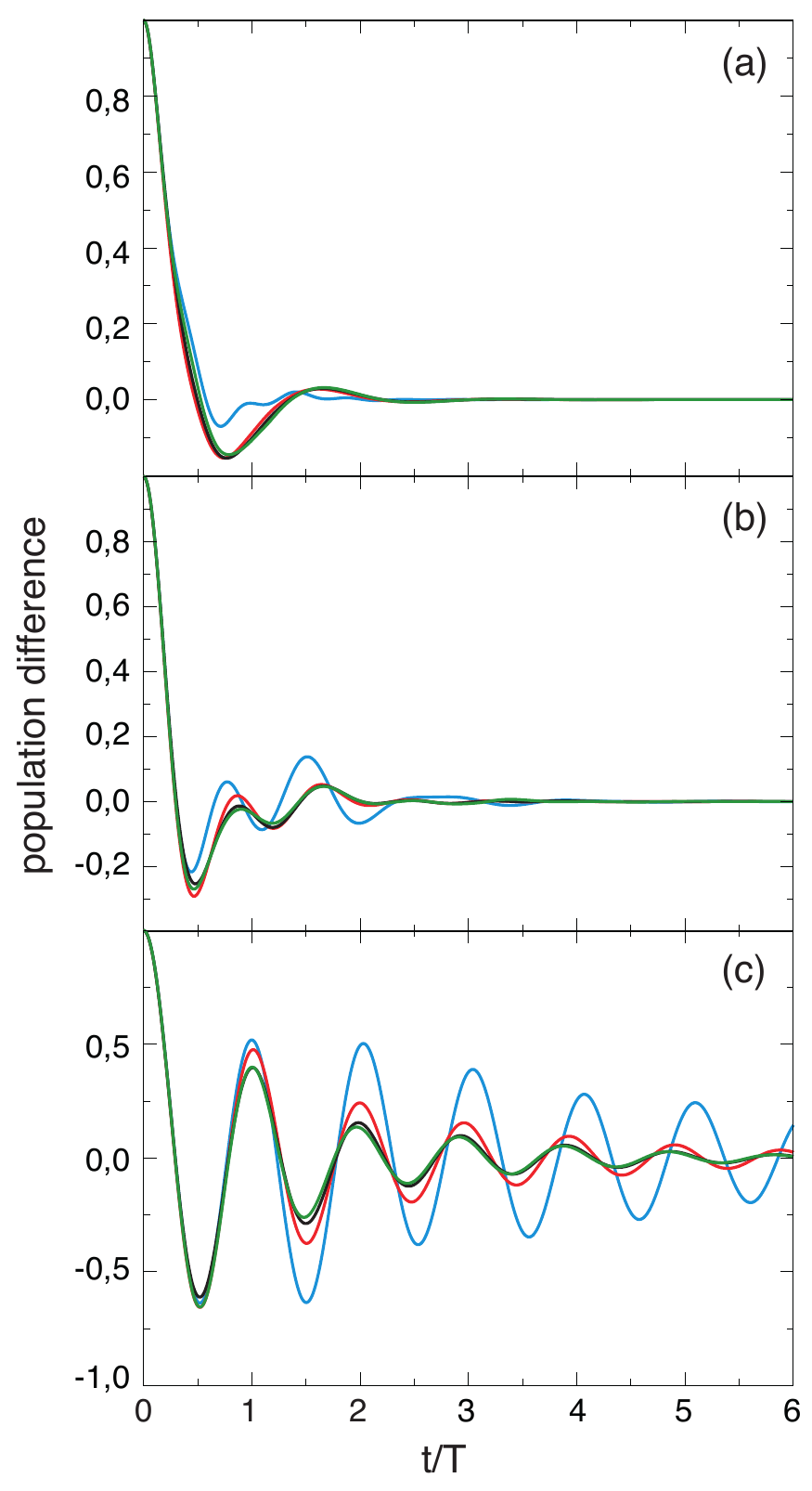}
\caption{Population difference dynamics of the homodimer model ($\Delta \epsilon=\unit[0]{cm^{-1}}$ coupled to a BO bath ($\lambda_{\rm BO}=\unit[50]{cm^{-1}}$,  $\gamma_{\rm BO}=\unit[50]{cm^{-1}}$). The parameters are (a) $J_{12}=\unit[50]{cm^{-1}}$ ($T=\unit[331]{fs}$), (b)  $J_{12}=\unit[100]{cm^{-1}}$ ($T=\unit[165]{fs}$), and (c)  $J_{12}=\unit[200]{cm^{-1}}$ ($T=\unit[83]{fs}$).  Color code: light blue -- exact HEOM, green -- non-equilibrium case Eq. \eqref{eq:Foerster_rate_homogeneous_term_line_shape_functions_noneq}, black -- equilibrium case Eq. \eqref{eq:Foerster_rate_homogeneous_term_line_shape_functions_std}, red -- fluctuation-only case Eq. \eqref{eq:Foerster_rate_homogeneous_term_matrix_elements_general_expression_noneq_no_int_pic}.}
\label{fig:5}
\end{figure}

So far we have focussed on a heterodimer. To complete the comparison between the different approximations we will next focus on the homodimer limit $\Delta \epsilon=0$. Figure \ref{fig:5} shows results for three different Coulomb couplings $J_{12}/\lambda_{\rm BO}=1$, $J_{12}/\lambda_{\rm BO}=2$, and $J_{12}/\lambda_{\rm BO}=4$ and for $\gamma_{\rm BO}=\unit[50]{cm^{-1}}$. The corresponding results for the heterodimer case are given in Fig. \ref{fig:3}. Interestingly, the results from the different second-order rate theories approximately agree with each other. This can be rationalized in terms of the F\"orster picture as follows. Considering forward donor to acceptor transfer, the vertical de-excitation transition at the donor (monomer 1) is between $\epsilon_1+\lambda_{\rm BO}$ and $\epsilon_1-\lambda_{\rm BO}$, depending on the progress of vibrational relaxation in the excited state. The acceptor excitation is at  $\epsilon_1-\Delta \epsilon+\lambda_{\rm BO}$ and should match the donor range for resonant transfer. If $\Delta \epsilon=0$, acceptor excitation is right at the center of the range of donor de-excitation, i.e. given a broad spectrum, transfer is possible no matter what the actual state of vibrational relaxation has been. This is in contrast to the case above where $\Delta \epsilon = \lambda_{\rm BO}$ allowing for resonance transfer in the lower part of the donor de-excitation range only. As a consequence the different  rate descriptions give in general different results. Of course, even for $\Delta \epsilon=0$ the different  descriptions might yield different population dynamics depending on the spectral overlap. Finally, we comment on the agreement with the exact reference shown in Fig. \ref{fig:5}. While for the cases $J_{12}/\lambda_{\rm BO}=1$ and  $J_{12}/\lambda_{\rm BO}=2$ it is surprisingly fair, for the strongest coupling, $J_{12}/\lambda_{\rm BO}=4$, only the first oscillation is reasonably reproduced.

\section{Beyond the Cumulant Approximation}
\label{sec:discussion}

In Sec. \ref{sec:cumulant} we have pointed out that  HEOM and second-order cumulant rates actually yield identical results for the present model. Therefore, one might ask what's the advantage of the present HEOM formulation and evaluation of second-order rates besides the proof-of-principle character. In fact the second-order HEOM rates are more general and can be obtained for cases beyond the limitations of the second-order cumulant approximation to the line shape functions. Two examples will be highlighted in the following. 

First, a variant of HEOM has been formulated for the case of Herzberg-Teller couplings \cite{SeMa18_CP_129}, i.e. for situations where the Coulomb coupling depends on vibrational coordinates.
The idea of this approach is to keep the hierarchy of ADOs as a data structure unchanged and to introduce the dependence of Coulomb coupling on vibrational coordinates, which is not taken into account in the standard formulation of the hierarchical equations, by relating the correlation function between such vibrational-coordinate-dependent coupling and the system bath coupling components to correlation functions of the latter. As the formulation of the HEOM method is based on the correlation functions of system-bath coupling components, additional terms arising from the correlation functions involving  Herzberg-Teller coupling can be added to the hierarchical equations with an appropriate scaling factor in the sense of a linear combination.
The same concept can, for example, also be applied for treatment of coupling between different bath components to account for a damping of an otherwise undamped oscillator due to coupling to a thermal bath. Such approach takes mechanistic aspects of the damping into account, instead of relying on a phenomenological damping constant as in the BO model.
While HEOM can be easily adjusted to include such effects, this is not the case in a treatment via cumulant expansion, which is aiming at a derivation of analytical expressions. Depending on the extent of vibrational-coordinate-dependence of the Coulomb couplings it is not possible anymore to obtain a sufficiently accurate second-order treatment when all components of the Coulomb coupling (also those with dependencies on vibrational coordinates) are taken into account in the interaction Hamiltonian. However, if the Coulomb coupling terms with dependencies on vibrational coordinates are taken into account in the reference Hamiltonian, the latter becomes non-diagonal and application of the cumulant expansion technique to obtain analytic rate expression in terms of line shape function is not practicable anymore. The calculation of the rates with HEOM can be applied in such cases without any adjustments.

Second, one might think of different separations into reference and interaction Hamiltonian. For instance, 
in the literature \cite{SiHa84_JCP_2615,LeMoCa12_JCP_204120,XuWaYaCa16_NJP_2016,Ki16_JCP_123,Ki16_JCP_70,SuFuIs16_JCP_204106} the polaron transformation has been applied for separation of reference and interaction Hamiltonian,
which can be formulated equivalently in the framework of a description with Liouville operators in HEOM space. Suppose that the polaron-transformed Hamiltonian $\hat{\bar {H}}=\hat{D}^{\dagger} \hat{H} \hat{D}$ is taken as the reference Hamiltonian, which because of the transformation does not exhibit displacements of bath coordinates in the electronic excited state manifold anymore. 
Even though the involvement of the polaron transformation in the separation of reference and interaction Hamiltonian suggests thermal equilibration in an initial excited-state population according to the considerations from Sec.~\ref{sec:extraction_second-order_rates}, this is not the case here because 
the interaction Hamiltonian is identified with the difference between untransformed and transformed Hamiltonian. Thus, the displacement of bath coordinates immediately after electronic excitation from the ground state to the excited state is actually contained in the interaction Hamiltonian and thus enters at the accuracy level of second-order cumulant expansion in the framework of second-order perturbative treatment in analogy to Eq.~(\ref{eq:QME_interaction_picture_tauPrime}). 

Specifically, the kernel in Eq. \eqref{eq:K2int} can be rewritten by starting from the decomposition of the total Liouvillian $\lsop{L}=\lsopBar{L}+\lsop{L}-\lsopBar{L}$ as follows
\begin{align}
\label{eq:QME_homogeneous_term_interaction_picture_with_pol_trans_separation_Href_Hint}
\lsop{K}(t,t')&= \lsop{P} \lsop{D}^\dagger e^{i\lsop{L}t} \lsop{D}(\lsop{L}-\lsopBar{L})\lsop{D}^\dagger e^{-i\lsop{L}t} \lsop{D}\lsop{Q} \lsop{D}^\dagger \nonumber\\
&\times e^{i\lsop{L}t'} \lsop{D}(\lsop{L}-\lsopBar{L})\lsop{D}^\dagger e^{-i\lsop{L}t'} \lsop{D}\lsop{P}	
\end{align}
Here, the projector is defined as in Eq. \eqref{eq:standard_projectors_HEOM_space}.

While the F\"{o}rster and modified Redfield approach in their respective bases imply a separation of diagonal and off-diagonal elements of the Hamiltonian by assignment to reference and interaction part, respectively, this is not the case if reference and interaction Hamiltonian are defined via a polaron transformation.  In fact, as the reference Hamiltonian in the present context contains both diagonal and off-diagonal elements, the off-diagonal elements, i.e.\ the excitonic coupling constants in the localized basis representation, become dependent on bath coordinates \cite{JaChReEa08_JCP_101104} and thus formally correspond to Herzberg-Teller couplings. The latter are straightforwardly treated using HEOM \cite{SeMa18_CP_129}.

\section{Conclusions} \label{sec:conclusions}

We studied the population dynamics of an excitonic dimer in the localized basis from integration of a QME with second-order transfer rates and compared the results with the exact population dynamics obtained from HEOM calculations. Starting from a generic formulation of the rate expressions in Liouville space, either HEOM propagations or the cumulant expansion technique can be applied to evaluate them. We pointed out the equivalence of both approaches for the considered model system and for the chosen way of calculating the second-order transfer rates. Different variants of the rate expressions can be formulated depending on the assumptions about the bath at the beginning of the transfer process, where it is either assumed to be non-equilibrated in the excited initial state of the transfer process due to vertical excitation from the electronic ground-state or to reside in thermal equilibrium due to preceding vibrational relaxation subsequent to electronic excitation. To treat the latter case with HEOM, we applied the concept of polaron transformation, which translates into hierarchical equations similar to those for time propagation in the context of HEOM. We identified the differences in the rate calculation under the assumptions of non-equilibrium and equilibrium initial bath at the level of the projection operator formalism in the framework of the solution of the Nakajima-Zwanzig equation. In this context we identified a further approach relying on the assumption of the commutability of projection operators and Liouville operators, which is only valid in the limiting case of thermal equilibration. As the evaluation via cumulant expansion indicates, this approach neglects vibrational relaxation, so that the transfer process is only driven by fluctuations in the respective description. Nevertheless, it can yield meaningful results in certain parameter regimes, particularly at the beginning of the transfer process. 

In our comparison of the population dynamics from second-order perturbative treatment and from non-perturbative treatment with HEOM, we started from the F\"{o}rster limit of small Coulomb coupling compared to the reorganization energy of the bath and increased the coupling to assess the applicability of the different ways of calculating second-order transfer rates beyond the F\"{o}rster regime, which turned out to depend on the further assumptions about the bath and the electronic system. In this respect we furthermore investigated the influence of the damping constant of overdamped and underdamped bath oscillators and of the energy gap between the excited states of the monomer units on the accuracy of the description. The results confirmed that the approach including only fluctuations does not reproduce an asymptotic population difference which appears for non-zero energy gap due to a preferred transfer direction determined by the resonance condition for the underlying combination of emission at the initial and absorption at the final site of the transfer process and by the thermal distribution of the involved bath phonons. For an energy gap of zero it turned out that the results from the different ways of second-order perturbative treatment become more similar, as there is no preferred transfer direction anymore. Furthermore, as expected, the non-equilibrium approach in general leads to a more reliable description than the equilibrium approach, at least at the beginning of the transfer process, whereas in the asymptotic limit the equilibrium approach leads to appropriate results and even outperforms the non-equilibrium approach in cases where higher-order effects lead to faster equilibration than the second-order non-equilibrium treatment predicts. For the calculation of the discussed results the rates were determined from expressions in terms of line shape functions obtained via cumulant expansion because their calculation using HEOM turned out to yield equivalent result, but with considerably larger numerical effort. Nevertheless, there are situations where the evaluation via cumulant expansion is not applicable and using HEOM ist an appropriate choice. We outlined an adjustment of rate calculation with separation of reference and interaction Hamiltonian via polaron transformation, where the interplay between Coulomb coupling and bath fluctuations is captured at the level of second-order perturbative treatment, for a description with HEOM. Such description can be accomplished by applying our concept of formulating the polaron transformation in the context of the hierarchical equations~\cite{seibt20}.

Although the present approach has been demonstrated for the case of a molecular dimer, application to transport in weakly coupled multichromophoric systems such as photosynthetic complexes~\cite{sener11_518} within the framework of rate equations is straightforward. Moreover, higher-order effects such as superexchange can be incorporated into the HEOM rates without much effort.


%
%



\end{document}